\documentclass[superscriptaddress,twocolumn,showpacs,showkeys,amsmath,amssymb,nofootinbib]{revtex4}
\usepackage{graphicx}
\usepackage{dcolumn}
\usepackage{bm}

\newcommand{\guig}{\textquotedblleft}
\newcommand{\guir}{\textquotedblright}

\begin{document}

\title{Probing neutron correlations through nuclear break-up}

\author{Marl\`ene Assi\'e}
\affiliation{Institut de Physique Nucl\'eaire, Universit\'e Paris-Sud-11-CNRS/IN2P3, 91406 Orsay, France}
\affiliation{GANIL, Bd Henri Becquerel, BP 55027, 14076 Caen Cedex 5, France}

\author{Denis Lacroix}
\affiliation{GANIL, Bd Henri Becquerel, BP 55027, 14076 Caen Cedex 5, France}


\begin{abstract}
The effect of initial correlations between nucleons on the nuclear break-up mechanism is studied. A quantum transport theory
which extends standard mean-field approach is developed to incorporate short range pairing correlation as well as direct 
nucleon-nucleon collisions. A time evolution of the nuclear break-up from a correlated system leading to the emission of two particles to the
continuum is performed. 
We show that initial correlations have strong influence on relative angles between particles emitted 
in coincidence. The present qualitative study indicates that nuclear break-up 
might be a tool to infer the residual interaction between nucleons in the nuclear medium.   
\end{abstract}
\maketitle

Nuclei are self-bound systems formed of fermions interacting through the strong nuclear interaction. 
While many facets of nuclei could be 
understood in term of independent particle motion, some aspects reveal internal 
correlations \cite{Boh75}. 
We consider here the so-called break-up process leading to the emission of nucleons to the 
continuum. Numerous dedicated models have been 
developed to account for this mechanism \cite{Alk03}. Among them, time dependent models
based on the independent particle hypothesis have been shown to provide a good description of the nuclear as well as 
Coulomb break-up \cite{dynall,Lac99}.
These approaches, by neglecting two-body
correlations could however not provide appropriate theories when 
two nucleons are emitted from the same nucleus \cite{Boa90,Lyn83,Mar00}. 
Interferometry measurements are being now analyzed using rather schematic models \cite{Boa90} and more elaborated 
theories.
are highly desirable.

The aim of the present work is twofold: (i) develop a microscopic quantum transport theory which incorporates effects beyond mean-field like pairing correlations and/or direct nucleon-nucleon scattering in the medium. (ii) present a qualitative study of nuclear break-up 
and show that this mechanism can be a tool of choice for the study of correlations in nuclei. 
Similar challenges to 
(i) are being now addressed in strongly correlated electronic systems using Time-Dependent Density Functional 
Theory (TDDFT) \cite{Mar06,App08}. The Energy Density Functional (EDF) \cite{Ben03} shares many aspects with DFT and 
is expected to provide 
a universal treatment of static and dynamical properties of nuclei \cite{Ben03,Sim08}. 
Current EDFs start from an effective interaction (of Skyrme or Gogny type) 
to provide an energy functional, denoted ${\cal E}(\rho)$, where $\rho$ is the one-body 
density matrix. Then, guided by the Hamiltonian case, equations of motion are        
written in terms of the one-body density evolution given by 
$i\hbar \partial_t \rho = [h[\rho],\rho]$, where $h[\rho]\equiv \partial {\cal E}(\rho) /{\partial \rho}$ denotes the mean-field Hamiltonian.
To account for the 
richness of phenomena in nuclear dynamics \cite{Lac04},
different extensions of mean-field have been proposed starting from  the evolution:
\begin{eqnarray}
i\hbar \frac{\partial \rho}{\partial t} = [h[\rho],\rho] + {\rm Tr}_2 [v^c_{12}, C_{12}] 
\label{eq:mfcor}
\end{eqnarray} 
where $v^c_{12}$ denotes the effective vertex in the correlation channel, ${\rm Tr}_{2}(.)$ is the partial trace on the second 
particle. $C_{12}$ denotes the two-body correlation defined from the two-body density $\rho_{12}$ as $C_{12} = \rho_{12} -\rho_1 \rho_2 (1-A_{12})$.
The indices refer to the particle on which the operator is applied 
(see for instance \cite{Lac04,Sim08}) while $A_{12}$ is the permutation operator. 
Eq. (\ref{eq:mfcor}) is generally complemented by the correlation evolution:
\begin{eqnarray}
i\hbar \frac{\partial C_{12}}{\partial t} = [h_1[\rho] + h_2[\rho] ,C_{12}] + B_{12} + P_{12} + H_{12},
\label{eq:corbph}
\end{eqnarray} 
where again the indices in $h$ refer to the particle to which the Hamiltonian is applied.
Expression of $B_{12}$ and $P_{12}$ and $H_{12}$, which can be found in \cite{Wan85,Lac04} 
are guided by the BBGKY hierarchy \cite{BBGKY}. These terms 
describe 
in-medium collisions, pairing and higher order effects respectively. 
{When three-body correlations are neglected, the above theory reduces to the so-called 
Time Dependent Density Matrix (TDDM) theory \cite{Wan85}.}
Coupled equations (\ref{eq:mfcor}) and (\ref{eq:corbph}) 
have been directly applied to giant resonances in ref. \cite{Tohall} and more 
recently to fusion in ref. \cite{Toh02}. However, applications are strongly constrained by the size of the 
two-body correlation matrix involved. 
In addition, similarly to ref. \cite{Pfi94}, we encountered difficulties to obtain numerical 
convergence towards a stable correlated system. Therefore, an appropriate approximation should be made
to render the TDDM theory more versatile. Keeping only $B_{12}$ and projecting out the effect of correlation onto the one-body evolution leads to the so-called Extended TDHF theory with a non-Markovian collision term \cite{Lac99b,Lac04}. Keeping only $P_{12}$ and assuming separable correlations leads to the Bogoliubov extension of the TDHF, i.e. TDHFB \cite{Toh04}. Both theories have been recently applied but require an 
\cite{Lac99b,Ave08}.

A different approximation is used here. We are interested in nuclei at low excitation where pairing 
plays an important role. Similarly to TDHFB, we group single-particles into pairs, denoted by 
$\{ \alpha, \bar \alpha \}$, where $| \bar \alpha  \rangle$ is initially the time-reversed state of $| \alpha  \rangle$.
We then assume that only components of ${v}^c_{12}$ and $C_{12}$ between such pairs are different from 
zero. 
This approximation, 
called hereafter TDDM$^{\rm P}$, leads to important simplifications: (i) the number of correlation
matrix components to be calculated is significantly reduced; (ii) the term $H_{12}$ cancels 
out. 
In the basis where $\rho$ is diagonal with occupation numbers $n_\alpha$, i.e. 
$\rho = \sum_\alpha| \alpha \rangle n_\alpha \langle \alpha |$, 
the evolution reduces to
\begin{eqnarray}
&& i\hbar \partial_t |\alpha \rangle =h[\rho] |\alpha\rangle ~~; ~~ \dot{n}_{\alpha}=\frac{2}{\hbar} \sum_\gamma \Im \left({\mathbf{V}}_{\alpha \gamma}
\mathbf{C}_{\gamma  \alpha }\right) \label{eq:rho_tddmp} \\
&& i \hbar\dot{\mathbf{C}}_{\alpha \beta}= 
{\mathbf{V}}_{\alpha\beta}((1-n_{\alpha})^{2} n_{\beta}^2-(1-n_{\beta})^{2} n_{\alpha}^2) \nonumber \\
&&+ \sum_\gamma {\mathbf{V}}_{\alpha \gamma }(1-2n_{\alpha})\mathbf{C}_{\gamma 
 \beta} -\sum_\gamma  {\mathbf{V}}_{\gamma \beta}(1-2n_{\beta})\mathbf{C}_{\alpha \gamma }\label{eq:c12_tddmp}
\end{eqnarray}
where $\mathbf{V}_{\alpha\beta} \equiv \langle\alpha \bar{\alpha}|v^c_{12}(1-A_{12})|\beta \bar{\beta} \rangle$, 
$\mathbf{C}_{\alpha\beta} \equiv \langle\alpha \bar{\alpha}|C_{12}|\beta \bar{\beta} \rangle$ and where 
the degeneracy of time-reversed states, i.e. $n_\alpha = n_{\bar \alpha}$ has been used. 

The TDDM$^{\rm P}$ incorporates correlations in the dynamics but also could be used to initialize a correlated system.
A correlated nucleus is obtained in two steps. First the ev8 code \cite{Bon04}
is used to obtain single particle states
which minimizes the Skyrme EDF using SIII interaction. Second, the TDHF3D code \cite{Kim97}, 
has been updated to incorporate both the evolution of correlation (Eq. (\ref{eq:corbph})) as well as its coupling 
to the one-body density (Eq. (\ref{eq:mfcor})). 
Equations are integrated using a second order Runge-Kutta method
where quantities on the right side of Eqs. (\ref{eq:rho_tddmp}-\ref{eq:c12_tddmp}) are estimated at time $t+\Delta t/2$ to perform 
the evolution from $t$ to $t+\Delta t$.  Note that this method insures the proper reorganization of the self-consistent mean-field when correlations built up. In particular, the mean-field polarization due to correlations is accounted for. 
{In this second step, following refs  \cite{Tohall,Pfi94}, we make use of the Gell-Mann-Low
adiabatic theorem \cite{Gel51} and switch on adiabatically the residual interaction by $v^c_{12}(t)\equiv v^c_{12}\left(1-e^{-t/\tau}\right)$. 
}
The residual interaction is set to \cite{Ben03,Ben05}: 
\begin{eqnarray}
%
v^c_{12}(\vec{r}_{1},\vec{r}_{2})&=&v_{0}
\left(1-\alpha \left[ \rho (\vec{R}) / \rho_{0}\right]^{\beta}\right)\delta(\vec{r}_{1}-\vec{r}_{2})
\label{eq:res}
\end{eqnarray}
where $\vec{R} \equiv (\vec{r}_{1} + \vec{r}_{2})/2$. The different parameters sets used here 
are given in table \ref{tab:forces} with $\rho_0=0.16$ fm$^{-3}$. 
\begin{table}[h!]
\begin{tabular}{p{2cm}p{2cm}p{2cm}p{2cm}}
\hline\noalign{\smallskip}
force & $v_{0}$ & $\alpha$ & $\beta$ \\
\noalign{\smallskip}\hline\noalign{\smallskip}
Attractive & -300& 1/2 & 1\\
Repulsive & +300 & 1/2 & 1\\
Volume  & -159.6 & 0 & -\\
Surface & -483.2 & 1 & 1\\
Mixed   & -248.5 & 1/2 & 1\\
\noalign{\smallskip}\hline
\end{tabular}
\caption{Parameters of the different residual interactions used in this work.}
\label{tab:forces}
\end{table}

\begin{figure}[hbtp]
\begin{center}
\includegraphics[height=3.5cm]{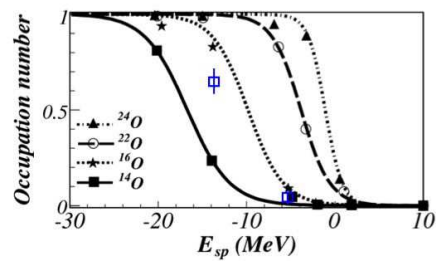}
\caption{Single-particle occupation numbers in oxygen isotopes as a function of single particle energies.
Lines represent Fermi functions fits. Open squares correspond to experimental occupation probabilities in $^{16}$O \cite{Lap93}.}
\label{fig:static}
\end{center}
\end{figure}

{Applications are performed 
in a three dimensional 
Cartesian mesh of size (80 fm)$^{3}$ with a step of 0.8 fm and a time step of 0.45 fm/c. 
}
For $\tau =300$ fm/c, a very good convergence of the adiabatic method, 
much better than for the full TDDM case \cite{Pfi94}, has been achieved. For this first application of TDDM$^{\rm P}$, we 
consider the isotopic oxygen chain with an $\alpha$ core while  
correlations build up between neutrons belonging to the $spd$ shells. 
After convergence, correlated systems have occupation numbers different from 0 and 1 and non-zero
correlation energy given by $E_{\rm corr} = \frac{1}{2}{\rm Tr}(v^c_{12} C_{12})$. Occupation numbers 
are displayed in Fig. \ref{fig:static}. TDDM$^{\rm P}$ goes beyond pure pairing 
theory like HFB due to the  
inclusion of two particles - two holes (2p-2h) terms. This is clearly illustrated by the doubly magic $^{16}$O nucleus 
where correlations do not cancel out and are in relatively good agreement with experimental observations of Refs. \cite{Lap93}. 
In table \ref{tab:pairing-gap}, an estimate of the pairing gap 
defined as \cite{Toh02a} $ {\Delta} \equiv 2 E_{\rm corr} / \sum_\alpha \sqrt{n_\alpha (1-n_\alpha)}$ 
is compared to the full TDDM and to the HFB cases. For the sake of comparison, the same core ($^{16}$O) and the same force as in ref. \cite{Toh02a} has been used.
{We see that our results are globally in agreement with the HFB and full TDDM 
results validating the approximation made in TDDM$^{\rm P}$.}
\begin{table}[h!]
\begin{tabular}{p{2cm}p{2cm}p{2cm}p{2cm}}
\hline\noalign{\smallskip}
 & TDDM \cite{Toh02a} & TDDM$^{P}$ & HFB \cite{Mat01}\\
\noalign{\smallskip}\hline\noalign{\smallskip}
$^{22}$O  & -3.1 MeV& -3.5 MeV& -3.3 MeV\\
$^{24}$O & -2.7 MeV & -3.1 MeV & -3.4 MeV\\
\noalign{\smallskip}\hline
\end{tabular}
\caption{Effective pairing gaps for $^{22-24}$O deduced respectively from TDDM \cite{Toh02a} and TDDM$^{\rm P}$ are compared 
to HFB results \cite{Mat01}.}
\label{tab:pairing-gap}
\end{table}

Correlated systems initialized with TDDM$^{\rm P}$ are then used to study the nuclear break-up 
leading to the  emission of two neutrons in coincidence. 
Intuitively, the two following scenarii have been proposed \cite{Ass08}: (i) if the two neutrons are initially close in position, 
both will feel the strong short range nuclear attraction of the reaction partner and will be emitted simultaneously at small relative angles (ii) if the two nucleons are far away in r-space, 
only one will undergo nuclear break-up. 
Then, the other nucleon might eventually be emitted isotropically from the 
daughter nucleus. Accordingly, large relative angles are expected between the two nucleons transmitted to the continuum in this sequential emission.     
To confirm this intuitive picture, the nuclear break-up dynamics 
is studied for an oxygen impinging on a $^{208}$Pb target.
\begin{figure}[hbtp]
\includegraphics[height=3cm,width=8.5cm]{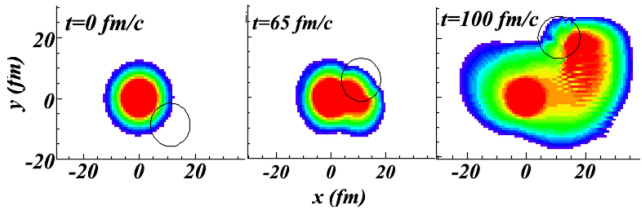}
\caption{(Color online) One body density for three different steps of the dynamical evolution for an $^{16}$O +$^{208}$Pb calculation at 40 A.MeV. The circle represents the $^{208}$Pb projectile.}
\label{fig:dyn}
\end{figure} 

{The correlated nucleus is first initialized at the center of a 3D mesh of size (80 fm)$^3$. 
Concerning the dynamical step, the collision is simulated treating the collision partner as a one-body 
time-dependent external perturbation, by replacing first equation in (\ref{eq:rho_tddmp}) by: 
\begin{eqnarray}
i\hbar \partial_t |\alpha(t) \rangle = \left\{h[\rho(t)] + V_{P}(\vec{r},t) \right\} | \alpha(t) \rangle
\end{eqnarray}  
where $V_{P}(\vec{r},t)$ stands for the projectile perturbation. Since we are 
considering here neutron emission, Coulomb effect is neglected and we assume 
that $V_P$ is a moving Woods-Saxon potential given by 
$V_{P}(\vec{r},t)= V_{0} / ( 1+\exp \{(|\vec{r}-\vec{r}_{0}(t)| -R_{\rm Pb} )/a \} )$ where $\vec{r}_{0}(t)$ corresponds 
to the lead center of mass position.
A simple straight line trajectory is used for $\vec{r}_{0}(t)$ corresponding to an impact parameter 
of 11 fm (grazing condition). The parameter ${R}_{\rm Pb}$ is set to 7.11 fm  
which corresponds to the $^{208}$Pb equivalent sharp radius while 
$V_{0}$=-50 MeV and $a=0.6$ fm. 
In the present 
application, we are interested in a specific reaction channel. If the full self-consistent 
calculation is made, many other elastic and inelastic channels would be populated and would render 
the extraction of the relevant information difficult. To focus on the effect of
correlation on the 
break-up channel we assume that only correlated neutrons are affected by the collisions partner. 
Therefore, particles in the core as well as protons are fixed during the evolution. In addition, we assume 
that occupation numbers and correlation matrix elements are kept equal to their initial values. 
The calculation is performed in the oxygen frame and 
an appropriate transformation is made to simulate an oxygen beam and 
obtain the proper angular distributions in the laboratory frame.
}
An example of one-body density evolution for the attractive force is shown
in Fig. \ref{fig:dyn}.
In Fig. \ref{fig:dyn}, three components can be distinguished: most of the neutrons remain in $^{16}$O, part of 
the nucleons (inside the circles) have been transferred while the rest is emitted to continuum. The last component corresponds 
to nuclear break-up emission and has already been understood in 
\cite{Lac99} and observed experimentally \cite{Sca98}.
\begin{figure}[hbtp]
\begin{center}
\begin{tabular}{cc}
\includegraphics[height=6cm]{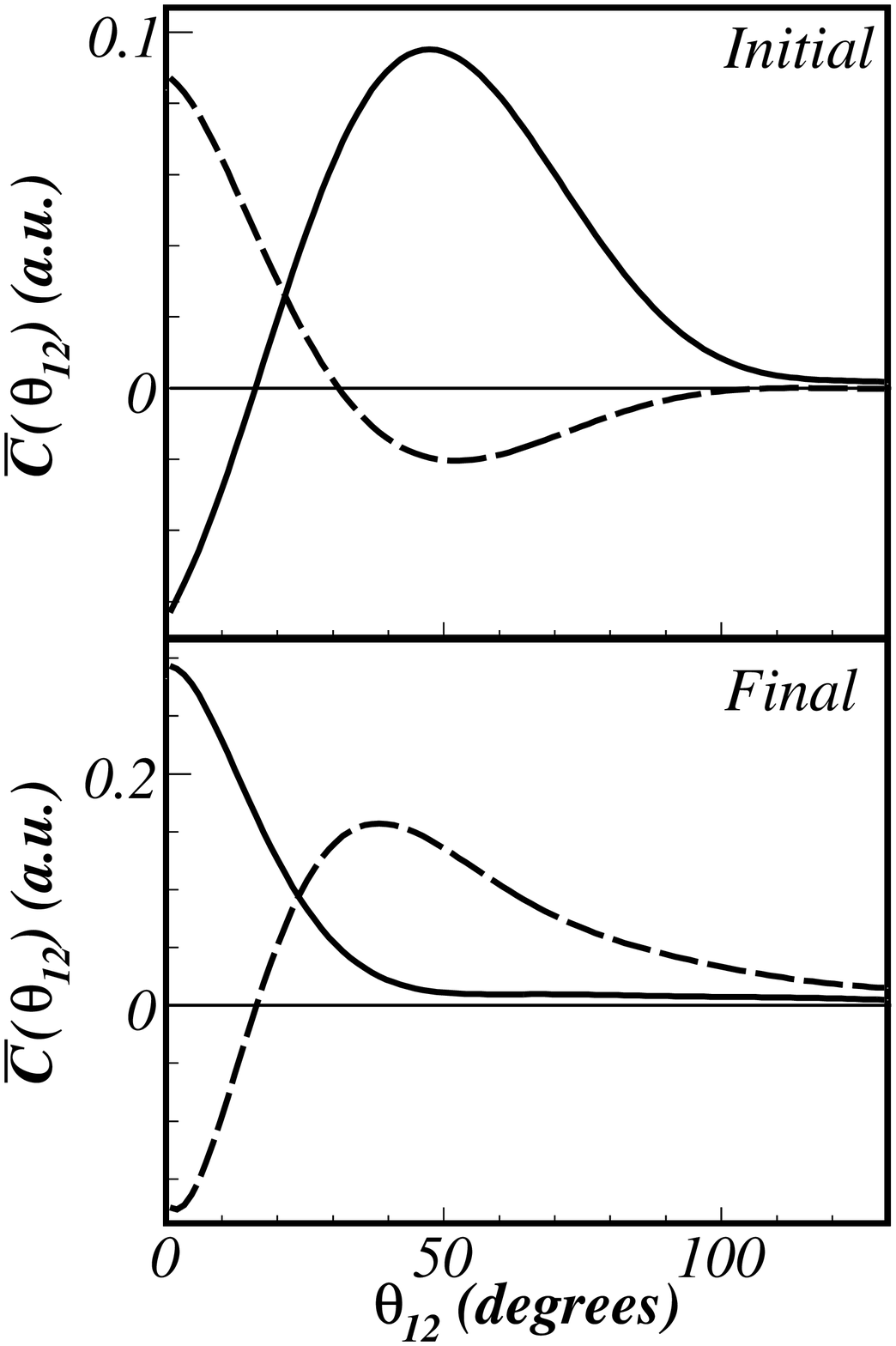} &
\includegraphics[height=6cm]{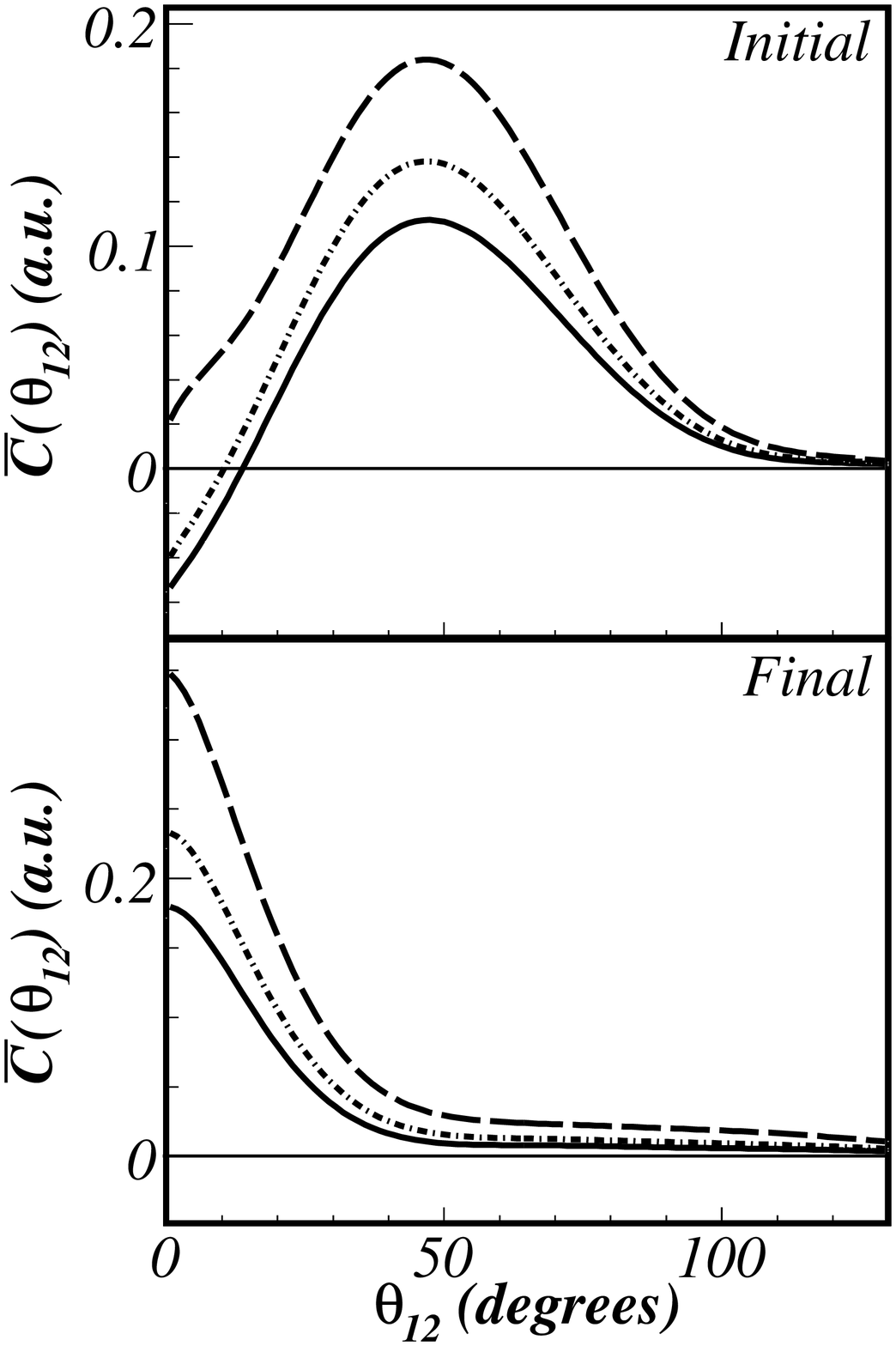}
\end{tabular}
\caption{
Left: Relative angle correlation between neutrons at initial (top) and final (bottom) time of the 
evolution for an $^{16}$O initialized with an attractive (full line)
or repulsive (dashed line) residual interaction.
Right: Initial (top) and final (bottom) relative angle correlation using the 
three different residual interactions : a \guig Volume\guir~(dashed line), 
\guig Surface\guir~ (full line) or \guig Mixed\guir~(dashed-dotted line) residual interaction.  All calculations are performed for an impact parameter of $b=11$ fm.}
\label{fig:pm}
\end{center}
\end{figure}
At the end of the evolution, nucleons from the inert core remaining in $^{16}$O or nucleons transferred to the collision partner are removed from 
the following analysis. The relative angles between two nucleons emitted in coincidence are then reconstructed 
from the correlation written in momentum space as: $C_{12}(\vec{p}_{1},\vec{p}_{2})=\sum_{\alpha \beta} \tilde{\phi}_{\alpha}(\vec{p}_{1})\tilde{\phi}_{\bar{\alpha}}(\vec{p}_{2})
 C_{\alpha\bar{\alpha}\beta\bar{\beta}}
\tilde{\phi}_{\beta}(\vec{p}_{1})\tilde{\phi}_{\bar{\beta}}(\vec{p}_{2})$,
where the $\tilde{\phi}_{\alpha}$ denotes the fraction of the final time single-particle states emitted to the continuum. 
The distribution of relative angles, $\bar{C}(\theta_{12})$, is then deduced by 
summing up contributions of all possible couples ($\vec{p}_{1}$,$\vec{p}_{2}$). 

To test the two scenarii, i.e. strong initial 
spatial correlation or anti-correlation, the 
\guig Attractive\guir~and \guig Repulsive\guir~interaction have been used.
{The \guig Repulsive\guir~interaction is used to 
mimic the \guig cigar\guir~like configurations, where the two correlated nucleons are well separated and on opposite side with respect 
to the core. Such a configuration is being now 
investigated experimentally using coulomb and/or nuclear break-up 
in lighter nuclei \cite{Mar00,Ass08}.}

The initial values of $\bar{C}(\theta_{12})$ are given in top-left
panel of Fig. \ref{fig:pm}. At initial time, the attractive force (close neutrons in r-space) 
leads to large relative momentum (and therefore large relative angles).
On the contrary, the repulsive force corresponds to small initial relative angles. In bottom-left panel, $\bar{C}(\theta_{12})$ 
is displayed at the final time of the evolution. Several important remarks could be drawn. First $\bar{C}(\theta_{12})$ is largely 
modified compared to the initial ones underlying the importance of dynamical effects. Therefore, experiments where two neutrons 
are detected in coincidence 
\cite{Mar00} could only be used if proper transport model are developed. Second, we see that our calculation indeed confirms 
the intuitive picture. Strong initial correlation in space leads to small relative angle emission (solid line) while 
for initially well separated nucleons, relative angles are much larger. 
Comparison of our model results with experiments requires to sum up the contribution 
of different impact parameters. It is worth mentioning that the shape of the angular correlation 
presented in Fig. \ref{fig:pm} does not change much as the impact parameter or Wood-Saxon parameters are modified. 
Therefore calculated cross-sections behave qualitatively as in  Fig. \ref{fig:pm}. In addition, experiments 
generally present the ratio $C^{exp}_{12} = P(\theta_{12})/P(\theta_1)P(\theta_2)$ where $P(\theta_{12})$ is the probability to emit two particles with a relative angle $\theta_{12}$. $\bar{C}(\theta_{12})$ calculated in this article is the difference between the correlated emission and the independent emission. It can be compared directly to the experiment by normalizing it to the independent emission and adding 1.
In conclusion, experiments dedicated to the study of 
two nucleon emission to the continuum due to the nuclear 
break-up used in parallel with dedicated transport models can provide a valuable method for the study 
of internal correlations in nuclei. 

We further investigated the sensitivity of relative angles with the initial correlations by comparing three realistic 
residual interactions called \guig Volume\guir, \guig Surface\guir~and \guig Mixed\guir~in table \ref{tab:forces} leading to the same scattering 
length \cite{Ben05}.
Corresponding initial and final angular correlations are respectively shown in top-right and bottom-right panel of Fig. \ref{fig:pm}.
Although all forces correspond to the scenario (i), sizeable differences are observed in the amplitude of $\bar{C}(\theta_{12})$.
Therefore, coincidence measurement of nucleon emission should be seriously considered in the near future as a tool to further 
constraint residual interaction used nowadays in EDF theories. It should however be kept in mind that a quantitative comparison with experiments requires a careful analysis of the present model parameters which is underway.

\begin{acknowledgments}
The authors thank Beno\^it Avez, Thomas Duguet and C\'edric Simenel for discussions.
We are especially grateful to  Jean-Antoine Scarpaci for his participation at different stages of this work. 
\end{acknowledgments}

\end{document}